\documentclass[journal]{IEEEtran}
\usepackage[T1]{fontenc}
\usepackage{amsmath,amssymb}
\usepackage{bm}
\usepackage{upgreek}
\usepackage{braket}
\usepackage{subcaption}
\usepackage{tikz}
\usepackage{here}

\ifCLASSINFOpdf

\else

\fi

\begin{document}

\title{Robust Position Sensing with Wave Fingerprints \\in Dynamic Complex Environments}

\author{Philipp~del~Hougne
\thanks{e-mail: philipp.delhougne@gmail.com}
}

\maketitle

\begin{abstract}
Irregular propagation environments with complex scattering effects challenge traditional ray-tracing-based localization. However, the environment's complexity enables solutions based on wave fingerprints (WFPs). Yet, since WFPs rely on the extreme sensitivity of the chaotic wave field to geometrical details, it is not clear how viable WFP techniques may be in a realistic dynamically evolving environment. Here, we reveal that environmental perturbations reduce both the diversity of the WFP dictionary and the effective signal-to-noise ratio (SNR), such that the amount of information that can be obtained per measurement is reduced. This unfavorable effect can, however, be fully compensated by taking more measurements. We show in simulations and experiments with a low-cost software-defined radio that WFP localization of non-cooperative objects is possible even when the scattering strength of the environmental perturbation significantly exceeds that of the object to be localized. Our results underline that diversity is only one important ingredient to achieve high sensing accuracy in compressed sensing, the other two being SNR and the choice of decoding method. We find that sacrificing diversity for SNR may be worthwhile and observe that artificial neural networks outperform traditional decoding methods in terms of the achieved sensing accuracy, especially at low SNR. 

\end{abstract}


\IEEEpeerreviewmaketitle

\section{Introduction}

\IEEEPARstart{P}{recise} position sensing is a highly-sought ability for countless context-aware devices in our modern life, including wireless communication with new-generation protocols relying on beam-forming, high-value asset tracking and customer analytics in retail, ambient-assisted living solutions for remote health care, intruder localization in classified facilities, and victim-detection technologies for first responders. Microwave-based sensing solutions are appealing due to their ability to operate through optically opaque materials or fog, their independence of external illumination and target color, limited potential privacy infringements and the non-ionizing nature of microwaves. Moreover, existing wireless infrastructure can often be leveraged, endowing it with a dual communication and sensing functionality.

Traditional microwave position sensing relies on ballistic wave propagation and leverages ray tracing approaches, the simplest example being triangulation. Unfortunately, the above-listed applications involve irregular propagation environments which give rise to significant multi-path effects. In some cases, the position to be identified may not even be within the sensor's line of sight but hidden around a corner. In such complex environments, a propagating wave front can get completely scrambled such that its angle or time of arrival cannot be used for position sensing with conventional ray-tracing analysis. 
Considerable research effort thus goes into overcoming the issues posed by multi-path effects, for instance, using distributed sensor networks encircling the region of interest in combination with a statistical analysis of shadowing effects and/or geometry-based environment models to account for reflections as virtual anchors~\cite{witrisal2016high,leitinger2019belief,mendrzik2019enabling,li2019massive,wymeersch2019radio}.

A completely different approach consists in embracing the complexity of the propagation medium as virtue rather than obstacle. An indoor environment is electrically large compared to the wavelength and can be characterized as ray-chaotic: the separation of two rays launched from the same location in slightly different directions will increase exponentially in time. A wave-chaotic field is extremely sensitive to both source location and the enclosure's geometry. Inspired by the quantum-mechanical concept of fidelity loss~\cite{kuhl2016microwave}, this sensitivity has been leveraged to distinguish nominally identical enclosures~\cite{hemmady2014apparatus}, to detect the presence or motion of small changes in the enclosure's geometry (without localizing them)~\cite{taddese2009sensor,del2018dynamic} as well as to quantify volume changing perturbations~\cite{taddese2013quantifying}. For the problem of position sensing, the wave-chaotic field's sensitivity implies that different positions are associated with distinguishable wave fields that can act like wave fingerprints (WFPs) for the positions. 

WFPing can be be applied to the localization of cooperative objects (emitting a beacon signal or equipped with a tag)~\cite{jin2008position,sen2012you,he2015wi,wu2015time,chen2016achieving,ing2005solid} as well as to non-cooperative objects (no compliance with localization task)~\cite{cohen2011subwavelength,xiao2013pilot,del2018precise}. While the former leverages the sensitivity of ray chaos to the source location, the latter leverages its sensitivity to geometrical perturbations. From the wave's point of view, different object positions inevitably correspond to different geometries of the propagation environment. To ensure the distinguishability of WFPs, the chaotic wave field must be probed in a number of "independent" ways. Traditionally, this is achieved using spatial or spectral diversity with a network of sensors or broadband measurements. A more recent alternative is to use configurational diversity by reprogramming the propagation environment with a "reconfigurable intelligent surface" (RIS). Using a programmable metasurface as RIS, Ref.~\cite{del2018precise} leveraged configurational diversity to localize multiple non-cooperative objects outside the line-of-sight with single-port single-frequency measurements.

With real-life applications in mind, a fundamental challenge for indoor localization with WFPs arises: how does one handle a dynamic evolution of the propagation environment independent of the objects of concern? Indeed, given the extreme sensitivity of the chaotic wave field to geometrical details, one could expect that a perturbation not related to the object to be localized alters the wave field to an extent that makes it irrecognizable in light of a previously established WFP dictionary. 

Here, we systematically study the impact of perturbations of the propagation environment on the localization accuracy, considering a frequency-diverse model system both in simulation and experiment. We investigate an interpretation of the perturber as effective source of noise and the extent to which the perturber affects the diversity of the WFP dictionary. We demonstrate that the reduction of the amount of information that can be obtained per measurement as the perturber size is increased can be compensated by taking more measurements, even in the regime where the perturber's scattering strength exceeds that of the object to be localized. Our results stress the importance of appreciating the information-theoretic encoding/decoding cycle of the sensing process in its entirety and reveal that machine-learning decoders outperform traditional decoding techniques especially in the low-SNR regime.

\section{Experimental Setup and WFP Formalism}

\begin{figure}
\centering
\includegraphics[width = \columnwidth]{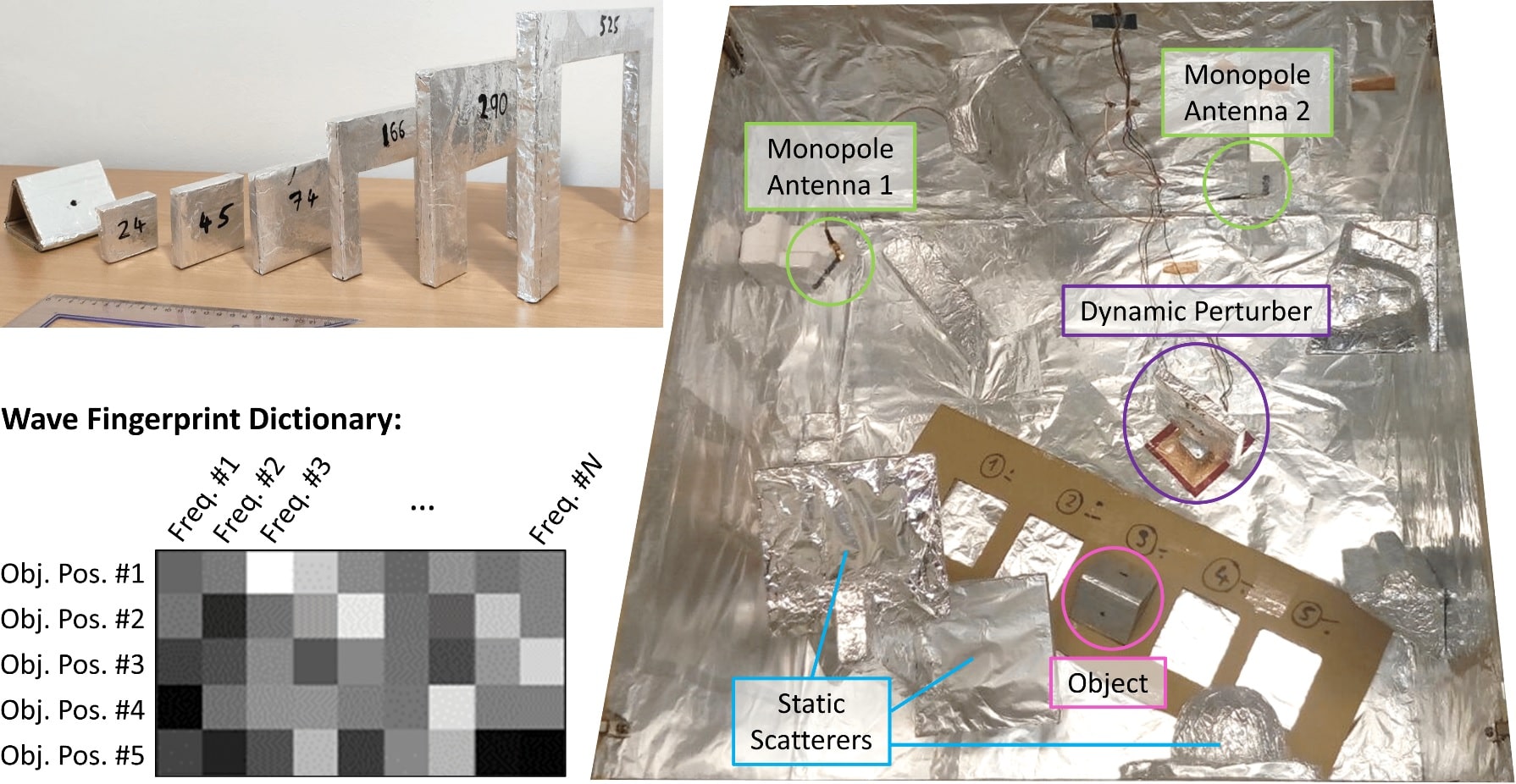}
\caption{Experimental setup (top wall removed to show interior). The triangular object to be localized (base $9 \times 9\ \mathrm{cm}^2$, height $6.5\ \mathrm{cm}$) is placed on one of $P=5$ predefined positions (here position \#3) inside a complex scattering environment. The transmission between two monopole antennas is measured with a LimeSDR Mini. The object to be localized is outside the antenna pair's line-of-sight. A dynamic perturber consists of a metallic object of variable size (here the third largest) mounted on a stepper motor. The top inset shows the different considered sizes of the dynamic perturber in comparison to the object size. The three largest perturbers are obtained by mounting a U-shaped extension on a smaller perturber, similar to the spirit of Matryoshka dolls. The bottom inset illustrates the WFP multiplexing mechanism.}
\end{figure} 

Our experimental setup is shown in Fig.~1: an object is located on one of $P=5$ possible predefined positions in an irregular metallic enclosure. $N=51$ complex-valued transmission measurements between two simple monopole antennas are taken in the interval $1\ \mathrm{GHz} < f < 2.58\ \mathrm{GHz}$ with a software-defined radio (SDR, LimeSDR Mini). Note that the predefined object positions are clearly outside the line-of-sight of the antenna pair. Dynamic perturbations of the propagation environment are introduced in our experiment with a metallic object of variable size mounted on a stepper motor which can place the object in an arbitrary angular orientation.

A measured transmission spectrum $S(f)$ can be decomposed into four contributions:

\begin{equation}
     S(f) = S_{cav}(f) + S_{obj}(f) + S_{pert}(f) + \mathcal{N}(f).
\end{equation}\label{Sdecomp}

\noindent $S_{pert}(f)$ accounts for rays that encountered the perturber, $S_{obj}(f)$ accounts for rays that encountered the object but not the perturber, $S_{cav}(f)$ accounts for rays that bounced around in the cavity without encountering object or perturber, and $\mathcal{N}(f)$ denotes the measurement noise. Given the chaotic nature of the complex scattering enclosure, it is customary to assume that real and imaginary components of the entries of the first three terms are drawn from zero-mean Gaussian distributions. The measurement noise is typically also zero-mean Gaussian. The decomposition in Eq.~1 has several subtleties. First, we note that if the perturber size is increased, more rays will encounter the perturber such that not only will the elements of $S_{pert}(f)$ be drawn from a distribution with larger standard deviation, but at the same time the standard deviation of the distributions of $S_{cav}(f)$ and $S_{obj}(f)$ will decrease. In other words, $S_{cav}(f)$ and $S_{obj}(f)$ are not independent of the perturbing object. Second, since all the terms are assumed to be drawn from zero-mean distributions, in principle one would expect that by averaging over an ensemble of realizations of the perturber one can estimate $S_{cav}(f) + S_{obj}(f)$ and by additionally averaging over an ensemble of object positions one can identify $S_{cav}(f)$. In practice, proper averaging requires a sufficient number of realizations and $P=5$ may be insufficient for averaging over an ensemble of object positions.

In Eq.~1, only the term $S_{obj}(f)$ encodes information about the object position. To determine a WFP in the presence of a perturber, we therefore average $S(f)$ over an ensemble of representative perturber realizations. Here, it is relatively easy to ensure that the ensemble is sufficiently large to estimate $S_{cav}(f) + S_{obj}(f)$ properly. We can then either define the WFP as being $S_{cav}(f) + S_{obj}(f)$ or we can intend to approximate $S_{obj}(f)$ with 
\begin{equation}
     S^{(2)}_{obj}(f)=S_{cav}(f) + S_{obj}(f) - \langle S_{cav}(f) + S_{obj}(f)\rangle_{obj}.
\end{equation}\label{Sdecomp}
\noindent We will consider both options below and see that, counter-intuitively, the former one can be advantageous in certain cases. Moreover, Eq.~1 naturally suggests to interpret the perturber as an effective source of noise. We can quantify the scattering strength of the perturber relative to that of the object via an effective perturber-induced SNR $\rho_p$. Ideally, to that end, we would define $\sigma_{s}$ and $\sigma_{n}$ to be the standard deviation of the distributions from which the entries of $S_{obj}(f)$ and $S_{pert}(f)$, respectively, are drawn, to define $\rho_p = \sigma_{s}^2/\sigma_{n}^2$. In practice, we do not know $S_{obj}(f)$. Depending on whether we choose to use $S_{cav}(f) + S_{obj}(f)$ or $S^{(2)}_{obj}(f)$ as WFP, we can define $\sigma^{(1)}_{s}$ and $\sigma^{(2)}_{s}$ to be the respective standard deviation, yielding $\rho^{(1)}_p$ and $\rho^{(2)}_p$. These effective SNRs quantify to what extent the perturber acts as noise on our chosen WFP, but do not directly reflect the ratio of scattering strengths of object and perturber.

The $P \times N $ WFP dictionary $\mathbf{H}$ merges the $P$ WFPs (each WFP is an $N$-element vector) into a single matrix. The WFP approach can then also be framed as a multiplexing problem $Y=\mathbf{H}X+\mathcal{N}$, where $X$ is a $1\times P$ vector identifying the object position, $Y$ is the complex-valued $1\times N$ measurement vector and $\mathcal{N}$ is a $1\times N$ noise vector.

\section{Information-Theoretic Perspective}

One prerequisite for successful WFPing is the diversity of $\mathbf{H}$. In our case, the complexity of the propagation environment naturally provides this diversity. The lower the correlations between different WFPs are, the better they can be distinguished. To get a quantitative grasp of the diversity of $\mathbf{H}$, it is instructive to consider its singular value (SV) decomposition: $\mathbf{H} = \mathbf{U} \mathbf{\Sigma} \mathbf{V}^T$, where $\mathbf{\Sigma} $ is a diagonal matrix whose $i$th entry is the $i$th SV $\sigma_i$ of $\mathbf{H}$. The flatter the SV spectrum is, the more diverse is $\mathbf{H}$. A convenient metric of diversity is the effective rank of $\mathbf{H}$ which is defined as $R_{\mathrm{eff}} = \mathrm{exp}\left( -\sum_{i=1}^n \tilde{\sigma}_i \mathrm{ln}(\tilde{\sigma}_i) \right)$, where $\tilde{\sigma}_i = \sigma_i / \sum_{i=1}^n \sigma_i$ and $n=\mathrm{min}(N,P)$~\cite{roy2007effective}.  
Note that only perfectly orthogonal channels with zero correlation yield $R_{\mathrm{eff}} = n$.

Unfortunately, much of the compressed sensing literature is exclusively focused on the diversity of $\mathbf{H}$ to understand the achievable performance. For instance, compression ratios are often provided without even indicating at what SNR they are valid. In principle, in the absence of any noise, the tiniest amount of diversity could be sufficient to ensure complete distinguishability even with $N=1$. Here, we argue that the achievable performance depends on the amount of (useful) information that can be extracted per measurement. In the physical layer, besides diversity the SNR is a second crucial ingredient. Moreover, high diversity and low SNR only ensure good performance if the deployed decoding method in the digital layer is capable of extracting much of the relevant encoded information from the measurement.

\begin{figure}[t]
\centering
\includegraphics[width = \columnwidth]{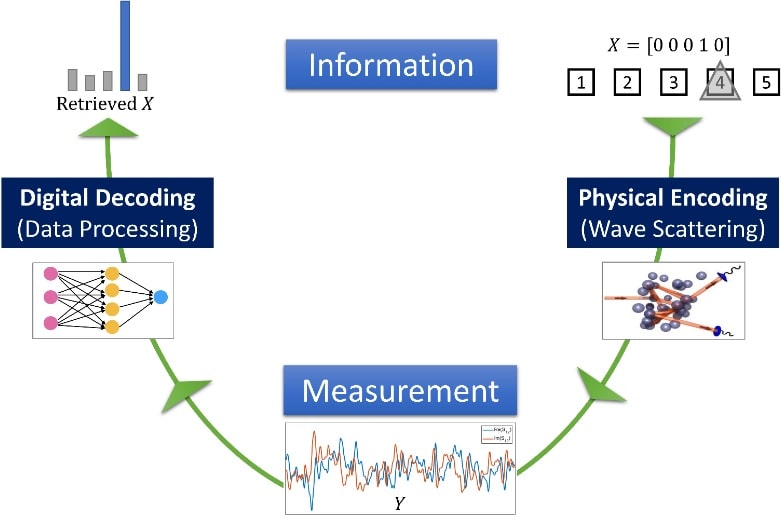}
\caption{Information about the object position is (inevitably) physically encoded in the measured data via wave scattering in the irregular propagation environment. Digital data processing then seeks to retrieve the information from the measurements.}
\end{figure} \label{figIT}

WFP-based sensing in its entirety as schematically summarized in Fig.~2 can be interpreted as a process consisting of physical encoding and digital decoding of information. Wave propagation through the complex scattering environment naturally (and inevitably) encodes information about the object position in measurements of the wave field. Data processing seeks to retrieve this information. Various decoding methods exist that we will compare later on:

(i) \textit{Correlation.} Identify which row of $\mathbf{H}$ has the highest correlations with $Y$. This procedure can be interpreted as "virtual time reversal"~\cite{ing2005solid}.

(ii) \textit{Inversion.} Compute an inverse of $\mathbf{H}$, for instance, via Tikhonov regularization, and identify the entry of $\mathbf{H}^{-1}Y$ with the largest magnitude. 

(iii) \textit{Optimized Inversion.} Use the result from (ii) as initial guess in a non-linear minimization of $|| Y-\mathbf{H}X ||$~\cite{redding2012using,redding2013compact}.

(iv) \textit{Learning.} Train an artificial neural network (ANN) to map $Y$ to the corresponding object position. ANN-based approaches have not been studied in the multiplexing literature to date. Besides their potential for superior decoding performance, inference is extremely fast. One forward pass through an ANN only requires a few matrix multiplications but no correlations, matrix inversions or nonlinear optimization routines.

From an information-theoretic perspective, it is important to understand fundamental bounds on the sensing performance.  
A simple bound to compute is the generalized Shannon capacity 
\begin{equation}
     C =  \sum_i \mathrm{log}_2 \left( 1+\frac{\rho}{P}\sigma_i \right)
\end{equation}\label{Sdecomp}
which has been mentioned on a few occasions in a sensing context~\cite{migliore2008electromagnetics,lorenzo2015single}. Nonetheless, the meaningfulness of $C$ for a specific sensing scheme is limited for two reasons. First, an ideal input distribution is assumed for $X$ but in reality all entries of $X$ are zero except for one which is unity. Second, an ideal decoding method is assumed. Below we will see  examples where a system with nominally lower $C$ nonetheless yields a higher sensing accuracy for certain decoding methods. 
It is thus essential to appreciate the sensing process in its entirety, including both encoding and decoding as illustrated in Fig.~2.

Having introduced the notion of diversity and SNR, we can now briefly comment on how faithfully the metallic enclosure in our experiment represents real-life scenarios. Without a doubt, certain cases like the inside of a vessel or a bank vault are very well represented. 
Other environments like the inside of a building are less reverberant than a metallic enclosure. Essentially, the quality factor of these "cavities" is lower. This implies more correlations within a fixed frequency interval of the transmission spectrum, as well as a lower SNR due to more attenuation. Both result in a decrease of the information that can be extracted per measurement; this effect can be compensated by taking more measurements, for instance, with a wider bandwidth. Nonetheless, from a fundamental perspective, the physics of an indoor system is entirely captured by our metallic enclosure. 
In scenarios with already existing wireless communication infrastructure, the beacon signals thereof could be used to implement position sensing with WFPs, 
saving energy and reducing the amount of electromagnetic radiation.

\section{Semi-Analytical Simulations}

To begin with, we consider a 2D version of our experiment simulated as a 2D system of coupled dipoles~\cite{orazbayev2020label} which contains all the essential physical ingredients to simulate wave propagation, reverberation and scattering in our experiment. 
These simulations offer an ideal platform to identify the effect of dynamic perturbations of the propagation environment on the sensing accuracy without any measurement noise or errors due to imperfect object positioning on the predefined positions, i.e. $\mathcal{N}(f)=0$. As shown in Fig.~3(a), a perturber of variable size with arbitrary orientation and location (within a specified area) simulates dynamic changes of the environment. Our simulation setup evaluates the transmission between an antenna pair at $25$ distinct frequencies. We use an ensemble of $150$ random perturber realizations (random orientation and random location of its center within the allowed area) to estimate $\mathbf{H}$, $R_{\mathrm{eff}}$ and $\rho_p$. The probability density function (PDF) of real and imaginary part of $S_{pert}$ is seen in Fig.~3(b-e) to be zero-mean single-peaked and tends towards a Gaussian distribution for larger perturbers. 

\begin{figure}
\centering
\includegraphics[width = \columnwidth]{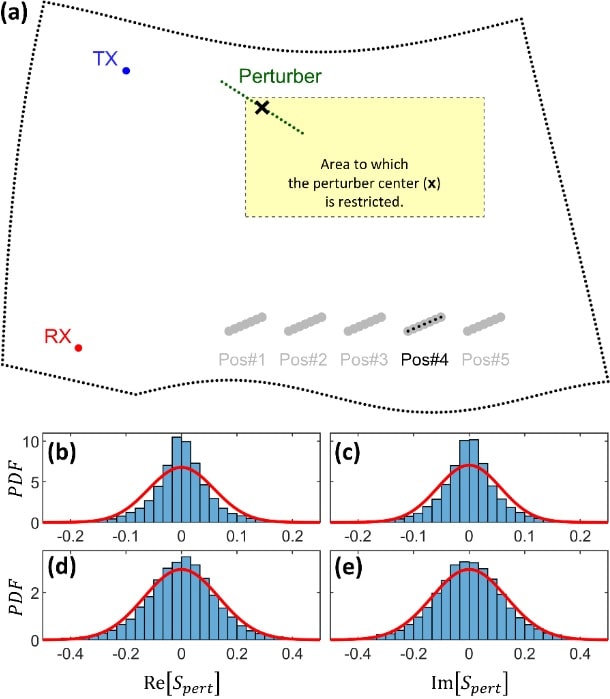}
\caption{(a) Setup of semi-analytical coupled-dipole 2D simulations. A line-like object is placed on one of $P=5$ predefined positions in an irregularly shaped enclosure (dipole fence) of dimensions on the order of $25\times15$ wavelengths. A line-like perturbing object with variable length is randomly rotated and located such that its center lies within the indicated area. The transmission between TX and RX is evaluated. See Ref.~\cite{orazbayev2020label} for technical details on the simulation method. (b-e) PDF of real and imaginary parts of $S_{pert}$ and a Gaussian fit are shown for the smallest and largest considered perturber size.}
\end{figure} \label{figCDCsetupX}

\subsection{Impact of Perturbation on Diversity and Effective SNR}

In Fig.~4 we contrast the use of $S_{cav}(f) + S_{obj}(f)$ or $S^{(2)}_{obj}(f)$ as WFP in terms of the resulting diversity ($R_{\mathrm{eff}}$), effective SNR ($\rho_p$) and sensing capacity ($C$). As we will see below, neither of these quantities is a reliable predictor of the sensing accuracy, since they do not take the decoding method into account. For the case of using $S_{cav}(f) + S_{obj}(f)$ as WFP, the observed trend is clear: as the perturber size increases, both $R_{\mathrm{eff}}$ and $\rho_p$ as well as $C$ decrease. While the impact on $\rho_p$ was clearly expected, the reduction of diversity is more subtle. It becomes intuitive by considering the extreme case in which the perturbation alters the entire enclosure. Then, averaging over realizations yields the result that would have been obtained in an anechoic environment such that no diversity thanks to wave chaos is left. 

\begin{figure}[h]
\centering
\includegraphics[width = \columnwidth]{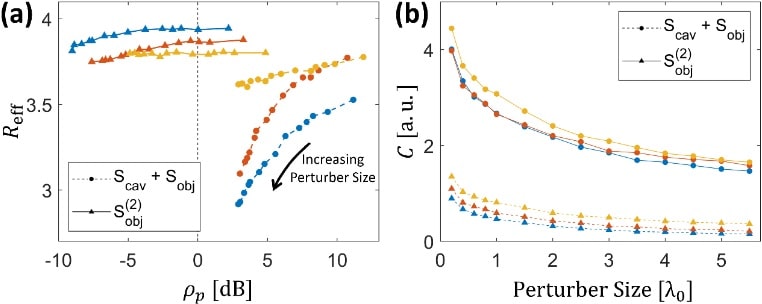}
\caption{(a) Effect of perturber size in the $R_{\mathrm{eff}}-\rho_p$ plane in the semi-analytical simulations. Curves for defining the WFP as $S_{cav}+S_{obj}$ or $S^{(2)}_{obj}$ are shown for three setups. All three setups are like the one in Fig.~3 but perturber area, predefined object positions and antenna positions are moved around. In all cases the objects are outside the antenna pair's line of sight. (b) Sensing capacity values corresponding to the data in (a). The results in this figure are obtained using all 25 frequency points. To ease comparison with Fig.~6, we normalized $\Sigma_{i=1}^n \sigma_i^2$ to unity; the SNR $\rho$ in Eq.~3 incorporates adverse effects on the dynamic range due to pathloss.}
\end{figure} \label{ReffrhoCDC}

Using $S_{obj}(f)$ as opposed to $S_{cav}(f) + S_{obj}(f)$ would certainly improve the diversity by removing unnecessary correlations (possibly at the expense of a better SNR such that the overall effect on capacity is unclear), but this is not possible in practice. Our closest option to that effect is to use $S^{(2)}_{obj}(f)$. Straight-forward simulations with random Gaussian matrices show that the effective rank of $S_{cav}(f) + S_{obj}(f)$ may exceed that of $S^{(2)}_{obj}(f)$ in cases where $P$ is small (preventing proper averaging over realizations of the object position) and where the ratio of the standard deviations of the distributions of $S_{obj}$ and $S_{cav}$ is large. Nonetheless, in our semi-analytical simulations, we observe in Fig.~4(a) a higher effective rank for $S^{(2)}_{obj}(f)$ than for $S_{cav}(f) + S_{obj}(f)$. Yet, since $\rho^{(2)}_p$ is substantially lower than $\rho^{(1)}_p$, the effect of using $S^{(2)}_{obj}(f)$ on the capacity is unfavorable. 

Complex scattering enclosures are often seen as random field  generators~\cite{corona1996reverberating}. $R_{\mathrm{eff}}$ is a measure of the number of independent samples and for $N\gg P$ one expects $R_{\mathrm{eff}} \rightarrow P$. Yet, in our simulations, $R_{\mathrm{eff}}$ saturates below $4$. This observation can be attributed to field correlations, here in the frequency domain, that prevent the field observables from being purely random variables~\cite{cozza2011skeptic}.

\subsection{Dependence of Sensing Accuracy on Perturber Size, Number of Measurements and Decoding Method}

The general trend is clear: the larger the perturbation, the less information can be extracted per measurement, as reflected by the sensing capacitance values plotted in Fig.~4(b). However, this decrease in information per measurement can be compensated with more measurements. At first sight, one may expect that such a compensation is only feasible as long as the object's scattering signature is stronger than the perturber's effect, i.e. for $\rho_p > 0\ \mathrm{dB}$. Our findings in Fig.~5, however, reveal that there is no abrupt phase change in the relation between achievable accuracy versus perturber size. Instead, using more measurements, successful position sensing is feasible at effective SNRs well below $0\ \mathrm{dB}$.

\begin{figure}[h]
\centering
\includegraphics[width = \columnwidth]{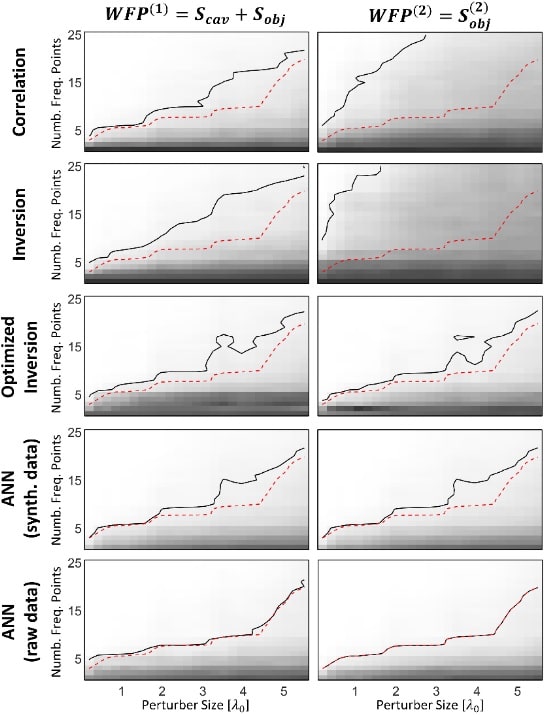}
\caption{Localization accuracy in semi-analytical simulations. The colorscale goes from 0 (black) to 1 (white). For each choice of WFP definition (columns) and decoding method (rows), the accuracy is plotted as a function of perturber size (horizontal axis) and the number of frequency points used to ink the WFP (vertical axis). ANN results are averaged over 20 training runs with randomly initialized weights; the standard deviation is below 2 \%. The black contour line corresponds to $95\ \%$ accuracy. To aid comparison, the red contour is the same on all subfigures.}
\end{figure} \label{AccCDC}

We systematically compare the previously outlined decoding methods for both choices of WFP. For the learning-based approach, we train an artificial neural network (ANN) consisting of two fully connected layers; the first layer consists of 256 neurons and is followed by a ReLu activation, the second layer consists of $P=5$ neurons and is followed by a SoftMax activation. Using more neurons or an additional layer does not appear to notably impact the results. We consider two possibilities to provide training data from which the ANN can learn to decode the measurements. The first option is to simply use the raw data from all the perturber realizations that we generated without a need for extracting $\mathbf{H}$ or other quantities. This brute force method may prove particularly useful in cases where measurements are restricted to intensity-only information which prevents averaging as simple means to extract $\mathbf{H}$, but this scenario is outside the scope of the present paper. Note that with this approach the WFPs are never explicitly evaluated, but only implicitly contained in the ANN weights. The second option is to synthesize training data with $Y=\mathbf{H}X+\mathcal{N}$ using the estimated $\mathbf{H}$ and generating $\mathcal{N}$ with entries drawn from a Gaussian distribution whose standard deviations match those of the distribution of $S_{pert}$ extracted from the data. This second method relies on our hypothesis that $S_{pert}$ is normally distributed and offers the possibility of generating a training dataset of unlimited size. In both cases we normalize the data (zero mean, unit variance) and use the Adam method for stochastic optimization (step size $10^{-3}$) to train the ANN weights.

In Fig.~5, we show how the achieved sensing accuracy depends on the perturber size and the number of measurements. We ensure that the spacing of the utilized frequency points is always the same and that they are always centered on the same frequency. For instance, for $N=7$ measurements we pick the central frequency point out of the 25 available ones as well as its three closest neighbours to the left and right. Our results are thus for one specific system realization which explains why the contours in Fig.~5 are not perfectly smooth. Several important observations and  conclusions follow from Fig.~5:

(i) WFP dictionaries with very different nominal sensing capacities can yield the same accuracy. This is the case for both ANN-based methods in which the accuracy is (almost) identical for $WFP^{(1)}$ and $WFP^{(2)}$.

(ii) The same WFP dictionary can yield very different accuracies depending on the decoding method. ANN-based decoders are seen to outperform correlation and inversion-based decoders.

(iii) The choice of WFP definition is irrelevant for the optimized inversion decoder as well as the ANN-based decoders. For correlation and inversion based decoders, however, using $WFP^{(1)}$ yields significantly better results.

(iv) Irrespective of the perturber size, we achieve an acceptable minimum accuracy (e.g. 95 \%). For larger perturbers, we need more measurements to compensate the reduction in the amount of information that can be extracted per measurement. Future information-theoretic work should seek to model the contour for a given accuracy in order to understand how the need for additional measurements scales with $\rho_p$.

(v) At low effective noise levels, some decoders achieve compression ratios above unity, that is they achieve accuracies $\geq 95\ \%$ to localize $P=5$ objects with $N<P$ measurements. For instance, the ANN (raw data) decoder with $WFP^{(2)}$ achieves 96 \% with $N=3$ at the lowest considered perturber size. However, as in any compressed sensing scenario, it is obvious that the compression ratio is heavily dependent on the noise level (here, the effective noise level due to the perturber size), the independence of different measurements (here, determined to a large extent by the interval between frequency points) and the decoding method (here, an ANN trained with raw data). Thus, a general claim of achieving a compression ratio above unity is not presented as key result of this work.

Overall, these results clearly demonstrate that it is fallacious to assume that the diversity or sensing capacity of $\mathbf{H}$ could be a reliable indicator of the sensing accuracy, hence the importance of considering the sensing process in its information-theoretic entirety as in Fig.~2.

\section{Experimental Results}

Having established an understanding of the perturber's effect under idealistic conditions in simulation, we now analyze the experimental data. Measurements with our SDR entail a few practical issues. First, there is a $\pm\pi$ uncertainty in measured phase values, originating from random phase jumps every time the Phase Locked Loop (PLL) is locked (e.g. to change the frequency). To obtain reliable data, we transform each measured complex value $z$ to $|z|\ \mathrm{exp}(2i\ \mathrm{mod}(\mathrm{arg}(z),\pi))$; the factor $2$ in the exponent ensures that the transformed variable's phase explores the entire $2\pi$ range. Second, the transmitted energy is clearly frequency-dependent, which can be caused by the frequency-dependent coupling of the monopole antennas to the cavity and/or frequency-dependent SDR components. The strong frequency-dependence means that we cannot simply model our variables as being drawn from a unique distribution, instead the distribution's standard deviation becomes frequency-dependent. To maintain the SDR's temperature constant throughout the experiment, we installed a simple CPU fan. We do not observe any significant amplitude or phase drifts over the course of the experiment.

We begin by quantifying two contributions to the $\mathcal{N}$ term in Eq.~1 that were not present in the simulations. First, we estimate the SNR due to measurement noise (by repeating the same measurement multiple times) as $\rho_1 = 25.5\ \mathrm{dB}$. Second, we estimate the SNR due to both measurement noise and imperfect positioning of the objects on the predefined locations (by repeating the same measurement multiple times after placing the object again on the same position) as  $\rho_2 = 15.8\ \mathrm{dB}$. 

\subsection{Impact of Perturbation on Diversity and Effective SNR}

\begin{figure}
\centering
\includegraphics[width = \columnwidth]{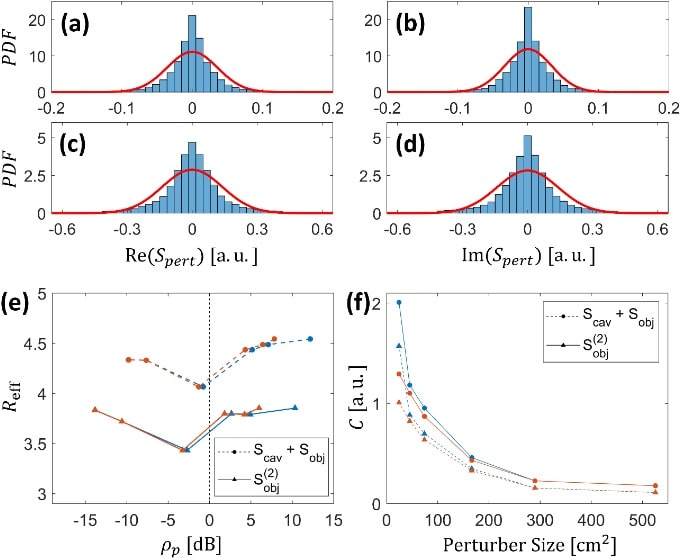}
\caption{(a-d) PDF of real and imaginary parts of $S_{pert}$ and a Gaussian fit are shown for the smallest and largest perturber size in the experiment.
(e) Effect of perturber size in the $R_{\mathrm{eff}}-\rho_p$ plane. 
The blue curves only consider perturber-induced effective noise, the red curves additionally account for measurement and positioning noise. 
(f) Normalized sensing capacity values corresponding to the data in (e). The results in this figure are obtained using all 51 frequency points.}
\end{figure} \label{figGNU_H_Analysis}

Based on 150 perturber realizations (random orientations) for each perturber size, in Fig.~6(a-d) we plot the PDFs of real and imaginary part of $S_{pert}$ for the smallest and largest perturber considered in our experiment. The zero-mean single-peaked distributions are identical for real and imaginary component but thinner than a Gaussian distribution. In Fig.~6(e) we plot $R_{\mathrm{eff}}(\mathbf{H})$ vs the effective SNR. Since $\mathcal{N}(f) \neq 0$ in the experiment, we plot two curves: the blue one only accounts for perturber-induced effective noise, the red one additionally accounts for measurement and positioning noise. The difference between these two curves is appreciable only for small perturber sizes since for larger perturbers $S_{pert}$ dominates over $\mathcal{N}$. Unlike in Fig.~4(a), using $S_{obj}^{(2)}$ lowers not only the effective SNR but also the effective rank. As in Fig.~4(b), we see in Fig.~6(f) that using $S_{obj}^{(2)}$ is unfavorable in terms of the (normalized) sensing capacity. The impact of $\mathcal{N}$ on $C$ is only noticeable for small perturbers.

\subsection{Dependence of Sensing Accuracy on Perturber Size, Number of Measurements and Decoding Method}

\begin{figure}
\centering
\includegraphics[width = \columnwidth]{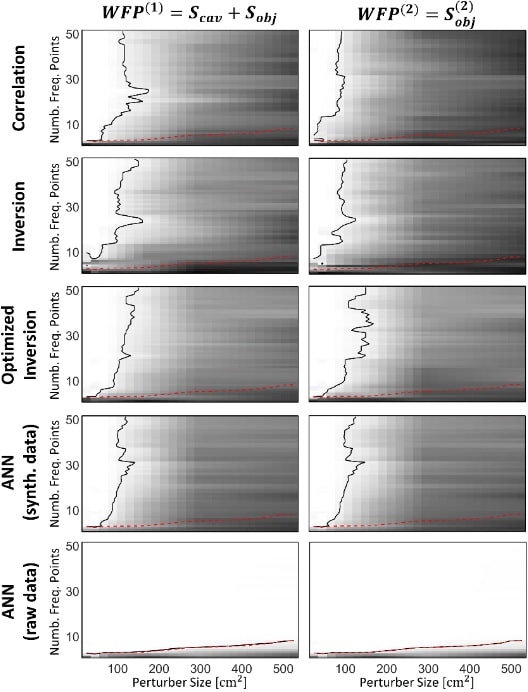}
\caption{Localization accuracy in experiments. The colorscale goes from 0 (black) to 1 (white). For each choice of WFP definition (columns) and decoding method (rows), the accuracy is plotted as a function of perturber size (horizontal axis) and the number of frequency points used to ink the WFP (vertical axis). ANN results are averaged over 20 training runs with randomly initialized weights; the standard deviation does not exceed 10 \% and 3 \% for ANNs trained with synthetic and raw data, respectively. The black contour line corresponds to $95\ \%$ accuracy. To aid comparison, the red contour is the same on all subfigures.}
\end{figure} \label{AccGNU}

In Fig.~7 we compare the achievable sensing accuracy in our experiment with the two considered definitions of the WFP and different decoding methods as a function of the perturber size and number of measured frequency points. The observations already made for the corresponding simulation results in Fig.~5 about the unsuitability of $R_{\mathrm{eff}}$ or $C$ to predict the sensing accuracy are confirmed once again by Fig.~7. The most notable difference to Fig.~5 is that except for the ANN trained with raw data all decoding methods fail to achieve at least 95 \% accuracy once the perturber's surface is larger than 200 $\mathrm{cm^2}$. We attribute this to the $\pm\pi$ phase uncertainty of our SDR which introduces errors in the estimation of $\mathbf{H}$. Since the ANN trained with raw data does not rely on calculating $\mathbf{H}$, it is not affected. Interestingly, we have thus a case in which it is better to feed the ANN raw data rather than to use physical insight to pre-process the ANN's training data. The ANN decoder trained with raw data is capable of achieving high sensing accuracies despite significant amounts of noise (the effective SNR is as low as -15~dB for the largest perturber, see Fig.~6(e)) and distorted data. 
Using the ANN decoder trained with raw data, we achieve 100 \% sensing accuracy with $N=3$, i.e. a compression ratio of $P/N=5/3>1$, for perturbers with a surface as large as $74\ \mathrm{cm}^2$. Again, we stress that the compression ratio depends on effective SNR, measurement independence and decoding method.

\section{Conclusion and Outlook}

From a practical point of view, our experiments, in combination with an ANN-based decoder, demonstrated the feasibility of precise position sensing with WFPs in dynamically evolving scattering enclosures using a low-cost and light-weight SDR. This capability is crucial to enable situational awareness in a plethora of emerging applications. Our technique does not rely on detailed knowledge about the environment's geometry and only requires a one-off calibration phase with multiple representative realizations of the dynamic perturbations that are expected during operation.
From a conceptual point of view, our work paves the way for a thorough information-theoretic analysis of sensing with WFPs. 
The dynamic perturber's unfavorable effect on diversity and effective SNR of the WFP dictionary, resulting in the acquisition of less useful information per measurement, can be fully compensated by taking more measurements -- even in the regime in which the perturber's scattering strength clearly exceeds that of the object to be localized.
We saw that the common practice in compressed sensing to only consider the diversity or capacity of $\mathbf{H}$ is insufficient to anticipate the achievable sensing accuracy. 
Our results are of very general nature: they can be applied to other types of wave phenomena (sound, light, ...) and are equally valid for WFPs established with other means such as using spatial or configurational degrees of freedom by employing a sensor network or a RIS~\cite{del2018precise}. 

The importance of seeing the entirety of the information-theoretic cycle points towards jointly optimizing encoding in a programmable propagation environment and ML-based decoding, as in the recently proposed "learned sensing" paradigm~\cite{del2020learned,li2020intelligent}. In contrast to compressed sensing which indiscriminately encodes all information, learned sensing seeks to encode only task-relevant information in the measurements. For position sensing, one could carefully select the frequencies at which measurements are taken (as opposed to linear spacing) and/or engineer the propagation environment with a RIS~\cite{del2020optimal}.

Looking ahead, it appears interesting to extend the present work (i) to scenarios with multiple objects to be localized, where neglected inter-object scattering is an additional effective source of noise~\cite{del2018precise}, (ii) to deeply sub-wavelength position sensing~\cite{cohen2011subwavelength}, and (iii) to more complex tasks like image transmission~\cite{li2018deep}.

In this work, the perturber was seen as an obstacle for our task to localize an object. In other contexts, the objective may be to characterize size and motion of a perturber. 
Diffuse wave spectroscopy~\cite{FishCounting,fishcountingPRL,CCSgeo,conti2006characterization} analyzes changes of the broadband impulse response over time to estimate the number or scattering cross section of objects moving through a complex medium. 
Our work has evidenced that the perturber's scattering strength can also be clearly related to the capacity of a multiplexing channel matrix averaged over different realizations of the perturber's position. Considering configuration-to-configuration multiplexing with two dynamic metasurface transceivers~\cite{sleasman2019implementation} may thus enable similar characterizations of a moving perturber with single-port single-frequency measurements~\cite{del2018dynamic}.

\section*{Acknowledgment}

The author acknowledges Michael del Hougne for help with the experimental work. The code for the semi-analytical model is based on Ref.~\cite{orazbayev2020label}. The code for controlling the SDR is based on Ref.~\cite{Lime}.

\ifCLASSOPTIONcaptionsoff
  \newpage
\fi

\bibliographystyle{IEEEtran}


\end{document}